\documentclass[aps,prl,twocolumn,superscriptaddress,showpacs,preprintnumbers]{revtex4}

\usepackage{graphicx}

\newcommand{\la}{\langle}
\newcommand{\ra}{\rangle}
\newcommand{\op}{{\cal O}}
\newcommand{\ndr}{{\rm NDR}}

\newcommand{\kslash}{k\kern-1ex /}
\newcommand{\pslash}{p\kern-1ex /}
\newcommand{\qslash}{q\kern-1ex /}
\newcommand{\lslash}{l\kern-1ex /}
\newcommand{\sslash}{s\kern-1ex /}
\newcommand{\Dslash}{{\cal D}\kern-1.5ex /}

\newcommand{\beqa}{\begin{eqnarray}}
\newcommand{\eeqa}{\end{eqnarray}}

\newcommand{\be}{\begin{equation}}
\newcommand{\ee}{\end{equation}}
\newcommand{\ben}{\begin{eqnarray}}
\newcommand{\een}{\end{eqnarray}}
\newcommand{\nn}{\nonumber}
\def\lsim{\raise0.3ex\hbox{$<$\kern-0.75em\raise-1.1ex\hbox{$\sim$}}}
\def\gsim{\raise0.3ex\hbox{$>$\kern-0.75em\raise-1.1ex\hbox{$\sim$}}}
\def\simgt{\rlap{\lower 3.5 pt\hbox{$\mathchar \sim$}}\raise 1pt \hbox {$>$}}
\def\simlt{\rlap{\lower 3.5 pt\hbox{$\mathchar \sim$}}\raise 1pt \hbox {$<$}}

\newcommand{\msbar}{{\overline {\rm MS}}}

\begin{document}

\newlength{\figwidth}
\setlength{\figwidth}{0.9\columnwidth}

\preprint{KEK-CP-148, UTHEP-485, UTCCP-P-148}

\title{Lattice QCD Calculation of the Proton Decay Matrix
Element\\ in the Continuum Limit
}

\newcommand{\Tsukuba}%
{Institute of Physics, University of Tsukuba,
 Tsukuba, Ibaraki 305-8571, Japan}

\newcommand{\RCCP}%
{Center for Computational Physics, University of Tsukuba,
 Tsukuba, Ibaraki 305-8577, Japan}

\newcommand{\ICRR}%
{Institute for Cosmic Ray Research, University of Tokyo,
 Kashiwa, Chiba 277-8582, Japan}

\newcommand{\KEK}%
{High Energy Accelerator Research Organization (KEK),
 Tsukuba, Ibaraki 305-0801, Japan}

\newcommand{\Hiroshima}%
{Department of Physics, Hiroshima University,
Higashi-Hiroshima, Hiroshima 739-8526, Japan}

\newcommand{\YITP}%
{Yukawa Institute for Theoretical Physics, Kyoto University,
 Kyoto 606-8502, Japan}

\author{N.~Tsutsui}
\affiliation{\KEK}

\author{S.~Aoki}
\affiliation{\Tsukuba}

\author{M.~Fukugita}
\affiliation{\ICRR}

\author{S.~Hashimoto}
\affiliation{\KEK}

\author{K-I.~Ishikawa}
\affiliation{\Hiroshima}

\author{N.~Ishizuka}
\affiliation{\Tsukuba}
\affiliation{\RCCP}

\author{Y.~Iwasaki}
\affiliation{\Tsukuba}
\affiliation{\RCCP}

\author{K.~Kanaya}
\affiliation{\Tsukuba}

\author{T.~Kaneko}
\affiliation{\KEK}

\author{Y.~Kuramashi}
\affiliation{\KEK}

\author{M.~Okawa}
\affiliation{\Hiroshima}

\author{T.~Onogi}
\affiliation{\YITP}

\author{Y.~Taniguchi}
\affiliation{\Tsukuba}

\author{A.~Ukawa}
\affiliation{\Tsukuba}
\affiliation{\RCCP}

\author{T.~Yoshi\'{e}}
\affiliation{\Tsukuba}
\affiliation{\RCCP}

\collaboration{CP-PACS and JLQCD Collaborations}
\noaffiliation

\date{\today}

\begin{abstract}
We present a quenched lattice QCD calculation of the  
$\alpha$ and $\beta$ parameters of the proton decay matrix element.
The simulation is carried out using the Wilson quark action
at three values of the lattice spacing in the range $a\approx$ 0.1$-$0.064 fm
to study the scaling violation effect.
We find only mild scaling violation 
when the lattice scale is determined by the nucleon mass.
We obtain in the continuum limit,
$|$$\alpha$($\ndr$,2GeV)$|$=0.0090(09)($^{+5}_{-19}$)GeV$^3$
and
$|$$\beta$($\ndr$,2GeV)$|$=0.0096(09)($^{+6}_{-20})$GeV$^3$
with $\alpha$ and $\beta$ in a relatively opposite sign, where
the first error is statistical and the second 
is due to the uncertainty in the determination of the physical scale.
\end{abstract}


\maketitle


Proton decay (or nucleon decay in general) is a characteristic consequence of
grand unified theories (GUTs)
because of the unification of quarks and leptons
into the same gauge multiplet.
However,
no clear evidence of such decay process has been observed up to now
in spite of continual experimental efforts over several decades.
Most recent experimental lower bound of the lifetime is given by
the Super-Kamiokande experiment:
$4.4\times 10^{33}$ years for $p\rightarrow e^++\pi^0$ mode
and $1.9\times 10^{33}$ years for $p\rightarrow {\bar \nu}+K^+$ mode
at 90\% confidence level\cite{sk}.
Although some naive GUT models is already ruled out by this experimental
bound, we still have several viable GUTs
which allow the longer proton lifetime at $O(10^{33-34})$\cite{guts}.
Further improvement of the experimental bound could give
strong constraints on these GUT models.

One of the main sources of uncertainties in the theoretical predictions is
the evaluation of the hadronic matrix elements for the nucleon decays
$\la PS | \op | N \ra$,
where $PS$ and $N$ stand for the pseudoscalar meson and the nucleon,
respectively, and $\op$ is the three-quark operator violating the baryon
number.
The conventional procedure of
estimating the hadronic matrix element is to 
invoke current algebra and PCAC and to
reduce the three-body matrix element into the two-body transition element
$\langle 0|{\cal O}|N\rangle$,
leaving aside the question as to the
validity of PCAC with a long extrapolation.
Varieties of models have been employed
to estimate this transition elements, but results vary by an order of
magnitude; see \cite{ellis}. 

A promising method to reduce the uncertainty is to resort to 
lattice QCD, which 
allows direct evaluation of non-perturbative effects and has been 
successfully used in giving various weak interaction
matrix elements. There are already a few calculations to evaluate the two-body 
transition element \cite{hara,bowler}, and even a few attempts 
to evaluate directly the three
body amplitude $\la PS | \op | N \ra$\cite{gavela,jlqcd_00}. 
Gavela et al.\cite{gavela}  argued that the three-body 
amplitude gives proton decay lifetime that differs largely 
from the one derived from a two-body calculation with the
use of PCAC. The JLQCD calculation \cite{jlqcd_00}, however,
showed that their results are due to a
neglect of one of the two form factors 
and that the three-body 
and two-body calculations yield the results that 
agree at a reasonable accuracy, say 20-30\%.

Lattice QCD calculations, being carried out today, however, contain a 
number of sources that lead to systematic errors, such as the quenching
approximation, finite lattice spacing, finite lattice size, chiral
extrapolation and so forth. Particularly worrisome are the finite
lattice spacing effects that could modify the continuum results
even by a factor 2 if scaling violation is substantial in the relevant
quantity. In fact, the recent
preliminary result of the RBC collaboration \cite{rbc_02} 
gives the matrix elements
that differs from those by JLQCD\cite{jlqcd_00} by $\approx$50\%,
which urges us to study the issue of systematic errors. 
 
In this paper we focus on the issue of the lattice spacing effects,
by carrying out simulations at three different values of bare coupling
constant, adopting the lattices that are large enough so that finite
lattice effects are negligible even for baryons, and borrowing the results
of a large-scale simulation of CP-PACS collaboration for quenched hadrons
\cite{cppacs}.
 We consider two-body matrix elements
\ben
 \la 0 |
 \epsilon_{ijk}
 ({u^i}^T CP_R d^j) P_L u^k
 | p(\vec{k}=\vec{0}) \ra &=&
 \alpha P_L u_p,\\
 \la 0 |
 \epsilon_{ijk}
 ({u^i}^T CP_L d^j) P_L u^k
 | p(\vec{k}=\vec{0}) \ra &=&
 \beta P_L u_p,
\een
expressed by $\alpha$ and $\beta$ parameters,
where
$i$, $j$ and $k$ are color indices, $C$ is the charge conjugation matrix,
$P_{R/L}$ is chiral projection operator and
$u_p$ denotes the proton spinor with the zero 
spatial momentum. We deal with the two-body matrix elements
in view of the feasibility on current computers, rather
than three-body matrix elements,
which need three-point correlators with finite spatial momenta injected
to disentangle the relevant and irrelevant form factors\cite{jlqcd_00}.


\begin{table*}
\caption{\label{tab:sim_para}
Simulation parameters and results. The lattice spacing $a$[fm]
in the third column is determined from $m_\rho$.
Results are given with three different input quantities for
the lattice spacing, \textit{i.e.} $m_N$, $m_\rho$, and
$f_\pi$.
}
\begin{ruledtabular}
\begin{tabular}{cccccccccc}
&&&\#conf.&\multicolumn{3}{c}{$|\alpha(\ndr,2{\rm GeV})|$[GeV$^3$]}
&\multicolumn{3}{c}{$|\beta(\ndr,2{\rm GeV})|$[GeV$^3$]}\\
 $\beta$ & $L^3\times T$ & $a$[fm] &
 this work/CP-PACS\protect{\cite{cppacs}}
& $m_N$ input & $m_\rho$ input & $f_\pi$ input & $m_N$ input & $m_\rho$ input & $f_\pi$ input \\
\hline
 5.90 & 32$^3\times$56 & 0.1020(8) & 300/800
& 0.01026(31) & 0.01265(43) & 0.01563(72) & 0.01064(32) & 0.01312(44) & 0.01621(76)\\
 6.10 & 40$^3\times$70 & 0.0777(7) & 200/600
& 0.01041(35) & 0.01152(41) & 0.01398(81) & 0.01092(37) & 0.01209(44) & 0.01467(85)\\
 6.25 & 48$^3\times$84 & 0.0642(7) & 140/420
& 0.00956(35) & 0.01012(45) & 0.01321(75) & 0.01004(35) & 0.01063(47) & 0.01388(78)\\
      &                & $a=0$         &
& 0.0090(09) & 0.0063(13) & 0.0092(22) & 0.0096(09) & 0.0069(14) & 0.0100(23)\\
\end{tabular}
\end{ruledtabular}
\end{table*}


The continuum operators relevant to the $\alpha$ and
$\beta$ parameters are connected with the lattice operators as
\ben
 \op_{R/L,L}^{\rm{cont}}(\mu) &=&
 Z(\alpha_s,\mu a) \op_{R/L,L}^{\rm{latt}}(a)
 +\frac{\alpha_s}{4\pi} Z_{\rm{mix}} \op_{L/R,L}^{\rm{latt}}(a)\nn\\
 &&\mp\frac{\alpha_s}{4\pi} Z^{\prime}_{\rm{mix}} \op_{\gamma_{\mu}L}^{\rm{latt}}(a),
\label{eq:zfactor}
\een
where
\begin{equation}
 \op_{R/L,L} =
 \epsilon_{ijk}
 ({u^i}^T CP_{R/L} d^j) P_L u^k,
\end{equation}
and the mixing operator
\begin{equation}
 \op_{\gamma_{\mu}L} =
 \epsilon_{ijk}
 ({u^i}^T C\gamma_{\mu}\gamma_5 d^j) P_L\gamma_{\mu} u^k
\end{equation}
appears due to explicit chiral symmetry breaking
of the Wilson quark action.
The renormalization constants $Z$, $Z_{\rm{mix}}$, and
$Z^{\prime}_{\rm{mix}}$ are evaluated perturbatively at
one-loop order\cite{pt_w,jlqcd_00}.
The continuum operators are 
defined in naive dimensional regularization (NDR) 
with the $\msbar$ subtraction scheme.
The matrix elements defined on the lattice are converted  to those 
in the continuum at $\mu=1/a$
and are evolved to  $\mu$=2GeV using
the two-loop renormalization group in the continuum\cite{na}.

To obtain the matrix elements, we consider the ratio
\ben
 R_{R/L}(t)&=&
 \frac{
 \sum_{\vec{x}}
 \la \op_{R/L,L}(\vec{x},t)
 \bar{J}^{\prime}_{p,s}(0)\ra
 }
 {
 \sum_{\vec{x}}\la J_{p,s}(\vec{x},t)\bar{J}^{\prime}_{p,s}(0)\ra
 }
 \sqrt{Z_p}\nn\\
&\stackrel{{\rm large}\;\;t}{\longrightarrow}&
      \la 0 | \epsilon_{ijk}({u^i}^T C P_{R/L} d^j) P_L u^k | p^{(s)} \ra,
\label{eq:rfunc}
\een
where $J_p(\vec{x},t)$ is a local sink operator 
for the proton with spin $s$ and
$\bar{J}^{\prime}_{p,s}(0)$ is the smeared source,
\ben
{ J}_{p,s}({\vec x},t)&=&
\epsilon_{ijk}
\left({u^i}^T(\vec{ x},t) C\gamma_5 d^j(\vec{ x},t)\right) u^k_s(\vec{ x},t),
\label{eq:local_p}\\
{ J}^\prime_{p,s}(t)&=&
\sum_{{\vec x},{\vec y},{\vec z}}\Psi({\vec x})\Psi({\vec y})\Psi({\vec z})\nn\\
&&\times\epsilon_{ijk}
\left({u^i}^T(\vec{x},t) C\gamma_5 d^j(\vec{y},t)\right) u^k_s(\vec{z},t)
\label{eq:smear_p}
\een
with the smearing function $\Psi$.
The factor $\sqrt{Z_p}$ defined by
\begin{equation}
 \la 0 | J_{p,s}({\vec 0},0) | p^{(s^\prime)}({\vec k}={\vec 0}) \ra =
 \sqrt{Z_p} u_s^{(s^\prime)} 
\end{equation}
is obtained from the
proton correlator with the local source and local sink.
It is recognized that the precise determination of $\sqrt{Z_p}$ is
not easy 
because of large statistical fluctuations of the local-local correlator
(see, {\it e.g.,} Fig.~16 of Ref.~\cite{cppacs}).
On the other hand, the ratio of two-point functions in
eq.(\ref{eq:rfunc}), calculated using 
the smeared-local proton correlator,
is determined well with small statistical errors.

Under this circumstance we calculate $\sqrt{Z_p}$ from 
the proton correlators generated in 
high statistics calculations of the quenched light hadron spectrum
performed by the CP-PACS collaboration\cite{cppacs}.
We then carry out new simulations with the same parameters to obtain 
the ratio of two-point functions 
including the mixing operator, for which we do not necessarily need 
very high statistics.
We attain a few percent statistical accuracy for the latter, 
while the overall accuracy is
still limited by the error of $\sqrt{Z_p}$.


\begin{figure}[t]
\includegraphics[width=\figwidth]{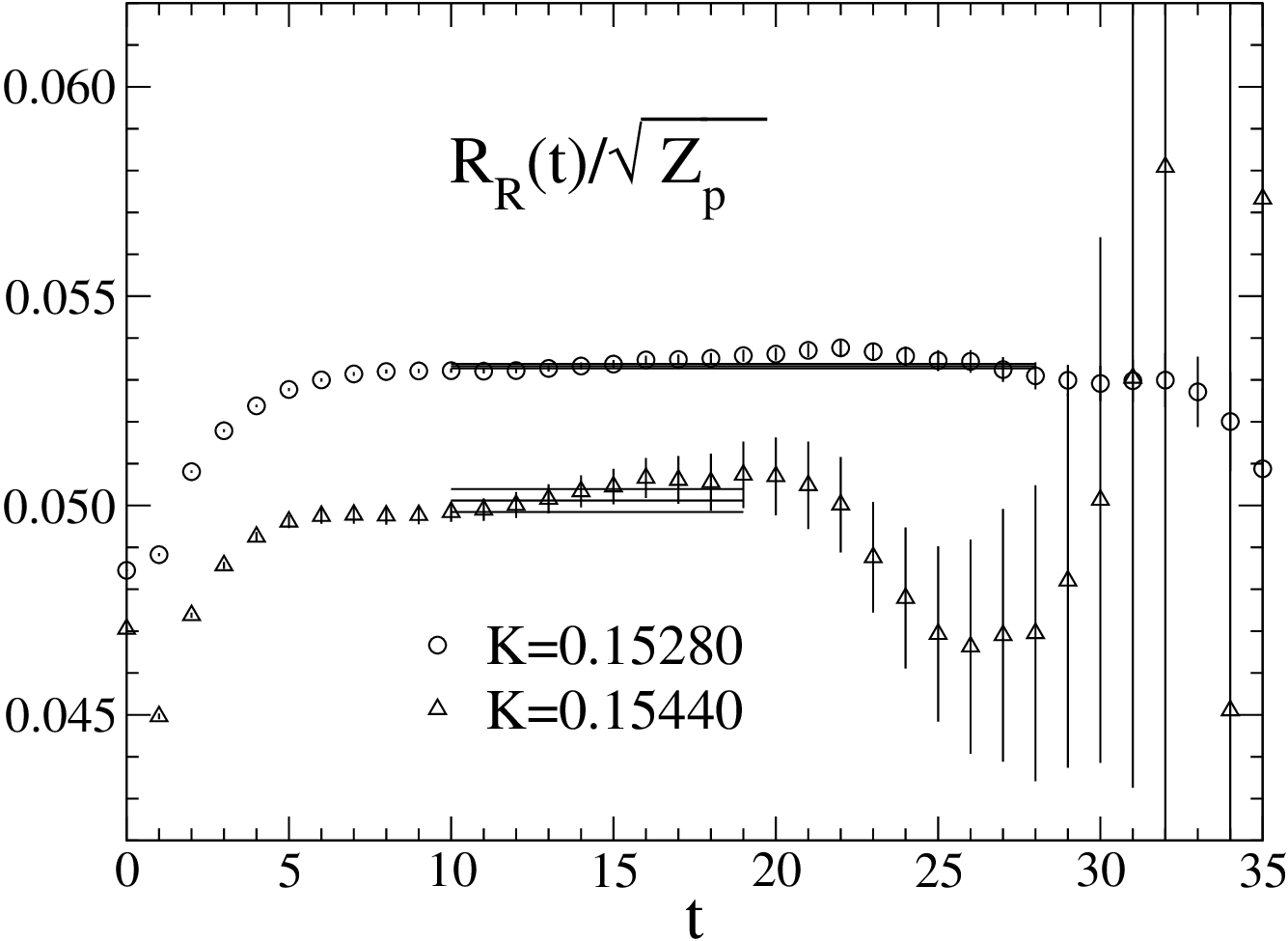}
\caption{\label{fig:rfunc}
Time dependence of $R_R(t)$ (the factor $\sqrt{Z_p}$ being removed)
for the heaviest (circle) and lightest (triangle)
quark masses at $\beta$=6.10.
}
\end{figure}


Our simulation generates quenched gauge configurations
at $\beta$=5.90, 6.10 and 6.25,
which correspond to lattice spacings in the range
$a\approx$ 0.1$-$0.064 fm
when determined from the $\rho$ meson mass 
$m_\rho$=0.7684 GeV.
The spatial lattice size is kept at about 3 fm
to avoid finite size effects.
We take four quark masses corresponding 
to $m_{PS}/m_{V}\approx$ 0.75$-$0.5 
for each $\beta$.
These parameters are the same as those of the CP-PACS
spectrum calculations \cite{cppacs}, except that we drop 
the finest lattice and the lightest quark mass 
at each $\beta$ for the computational cost.
The simulation parameters are presented in Table~\ref{tab:sim_para}.
The number of configurations in our simulation is about 1/3 
that of the CP-PACS calculation. 
We employ for $\Psi$ in eq.(\ref{eq:smear_p}) 
the pion quark wave function, which is  
measured for each hopping parameter 
on 30 gauge configurations fixed to the
Coulomb gauge except for the $t$=0 time slice 
where the wall source is placed\cite{wogf}.

\begin{figure}[t]
\includegraphics[width=\figwidth]{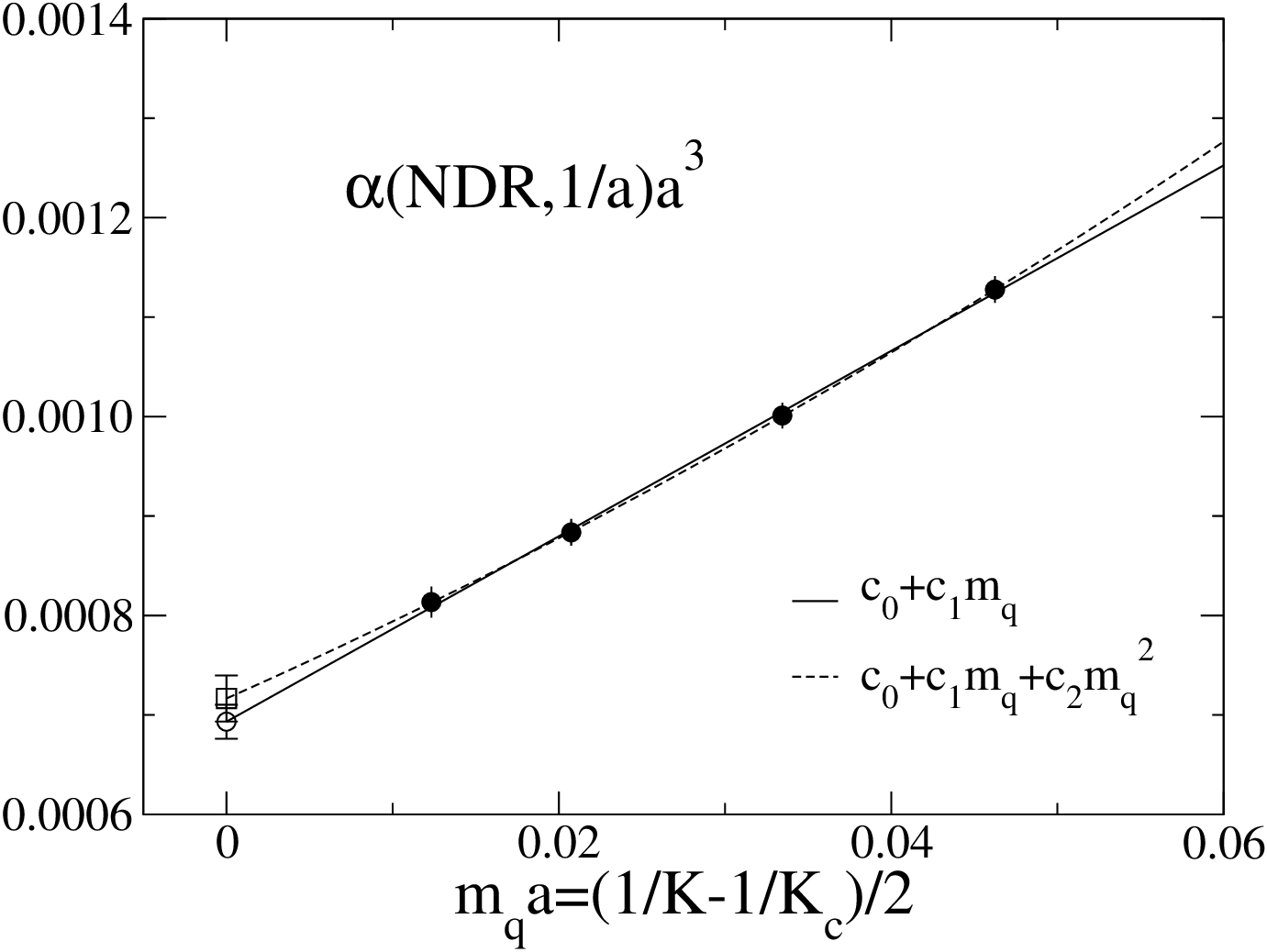}
\caption{\label{fig:mqdep}
Chiral extrapolation of the $\alpha$ parameter with linear (solid) and
quadratic (dotted) functions at $\beta$=6.10.}
\end{figure}

To estimate $\sqrt{Z_p}$, we fit the smeared-local proton correlator
to a single exponential $Z_p^\prime {\rm exp}(-m_p t)$,
and then fit the local-local proton correlator  
to $Z_p {\rm exp}(-m_p t)$ with $m_p$ fixed to the value determined
from the smeared-local correlator, which is borrowed from
the CP-PACS simulation\cite{cppacs}.
Figure~\ref{fig:rfunc} shows the ratio 
of two-point functions in eq.(\ref{eq:rfunc}) with the $\sqrt{Z_p}$
factor removed 
for the heaviest($K=0.15280$) and the
lightest($K=0.15440$) quark masses at $\beta=6.10$. The horizontal lines
represent the fits together with one standard deviation errors,
which are smaller than 1\%.
In Fig.~\ref{fig:mqdep} we plot
the quark mass dependence of the $\alpha$ parameter at $\beta=6.10$,
which is well described by a linear function.
A similar quark mass dependence is observed for the $\beta$ parameter. 
We find that linear plus quadratic extrapolations yield 
results consistent within error bars in the chiral limit at all 
lattice spacings.
The $\alpha$ and $\beta$ parameters
in the chiral limit obtained by linear extrapolations
are summarized in 
Table~\ref{tab:sim_para}. 
The errors are at most a few percent.
The contribution of the mixing operator 
in eq.(\ref{eq:zfactor}) is smaller than 10\%.

\begin{figure}[t]
\includegraphics[width=\figwidth]{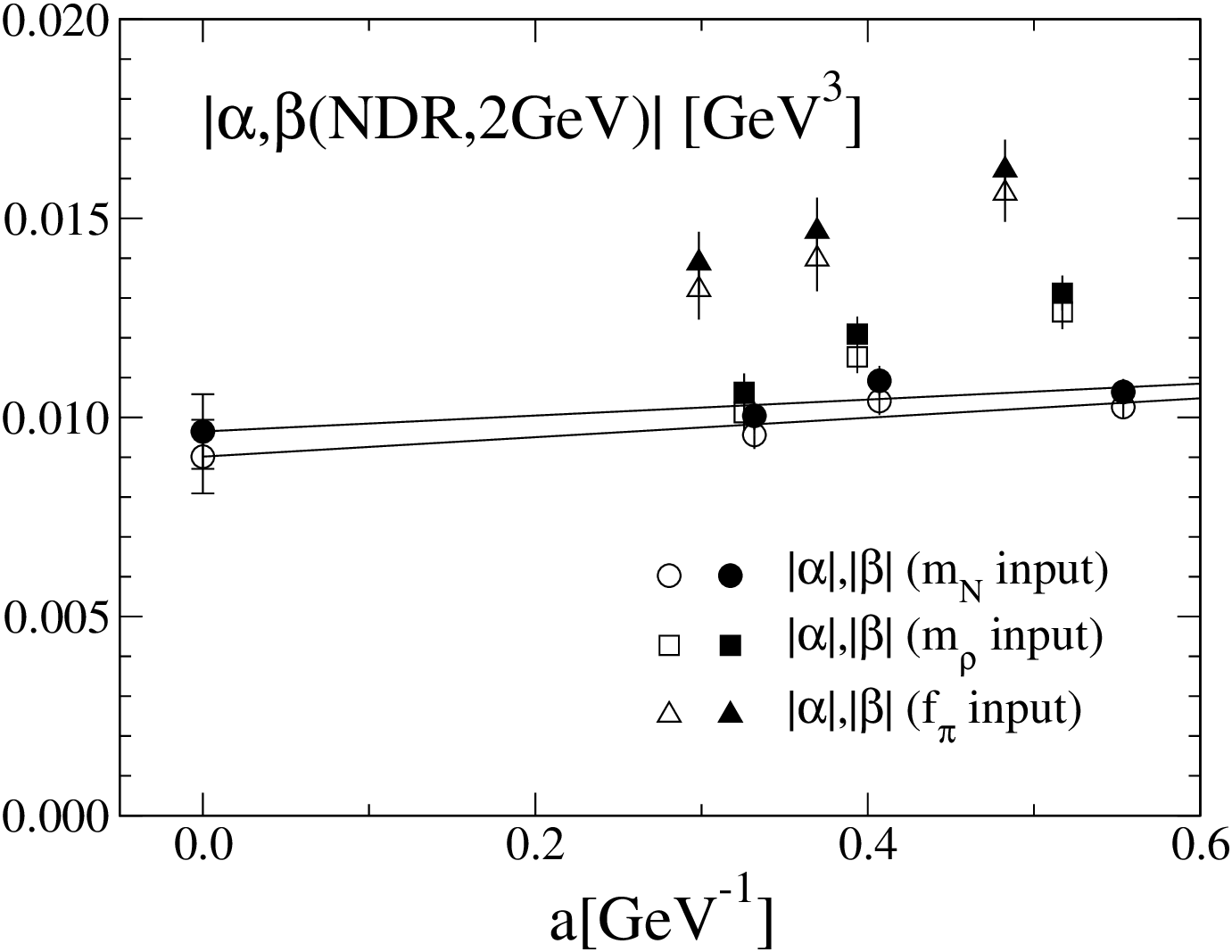}
\caption{\label{fig:adep}
Continuum extrapolation of 
$\alpha$ (open) and $\beta$ (filled) parameters.
The lattice scale is determined from $m_N$ (circle),
$m_\rho$ (square) and $f_\pi$ (triangle)
at each $\beta$.
The errors in the continuum limit are statistical only.
}
\end{figure}

We present in Fig.~\ref{fig:adep} $\alpha$ and $\beta$ 
in physical units as a function of $a$.
We examine three choices, the nucleon mass $m_N$, 
the $\rho$ mass $m_\rho$, and the pion decay constant $f_\pi$
to determine the physical scale of the lattice. We take these physical
parameters given by CP-PACS~\cite{cppacs}, because $\alpha$ and
$\beta$ have dimension 3 and the error of mass scale 
is magnified by a factor of 3, so that high statistics
results are essential.
Figure~\ref{fig:adep} indicates that scaling violation $\alpha$ 
and $\beta$ is minimized if nucleon mass is used as input.
The use of mesonic quantities, $m_\rho$ or $f_\pi$, on the other hand, 
leads to substantial scaling violation. 
A simple linear extrapolation to the continuum limit results in
$\alpha$ and $\beta$ that vary up to 30\% depending on the input
physical scale as found in Table~\ref{tab:sim_para}.
 
We adopt the $\alpha$ and $\beta$, extrapolated to $a=0$, using
the $m_N$ input as our central value, since small scaling violation
would minimize the error associated with the continuum extrapolation,  
and include the uncertainty in the physical scale as systematic error.
We obtain
\ben
 |\alpha(\ndr,2\textrm{GeV})|&=&0.0090(09)
 \left(^{+5}_{-19}\right)
 \textrm{GeV}^3,\\
 |\beta(\ndr,2\textrm{GeV})|&=&0.0096(09)
 \left(^{+6}_{-20}\right)
 \textrm{GeV}^3,
\een
where the first error is statistical and the second one is systematic.
Since the CP-PACS spectrum calculation\cite{cppacs} is superior to this work
in controlling the systematic errors using  
finer lattices and lighter quark masses than this simulation,
we estimate the ambiguity due to scale setting
from their results of quenched light hadron mass spectrum. 
They show that the values of $m_\rho$ and $f_\pi$ in quenched QCD deviate 
from the experiment by $+7$\% and $-2$\% 
respectively in the continuum limit,
once we set the lattice spacing by $m_N$.
The errors of mass scale in $\alpha$ and $\beta$ are magnified by a
factor of 3 and found to be comparable with the variation of the results
at the continuum limit in Table~\ref{tab:sim_para}.
This implies that the systematics from the physical scale dependence
are mostly ascribed to the quenching effects.
We note that the sign of $\alpha$ and $\beta$ are relatively opposite,
while the overall sign is a convention. 
Our results are about 3 times larger than
 the smallest estimate among
various QCD model predictions,
$|\alpha|=|\beta|=0.003$ GeV$^3$\cite{ab_min}, 
which is often used
in phenomenology of GUTs to derive
``conservative'' estimates of proton lifetime.


Our present $\alpha$ and $\beta$ are smaller than the previous results
using the same gauge and quark actions, 
$|\alpha(\ndr,1/a)|=0.015(1)$ GeV$^3$ and 
$|\beta(\ndr,1/a)|=0.014(1)$ GeV$^3$ at $1/a=2.30(4)$ GeV \cite{jlqcd_00},
beyond what is expected from scaling violation obtained in this work.
We suspect that $\sqrt{Z_p}$ and the lattice scale determined from 
$m_\rho$ are overestimated while their errors are not properly estimated, 
probably due to large fluctuations in
$\sqrt{Z_p}$ for which only 100 configurations were used. 
We also compare our $\alpha$ and $\beta$ with those
of the preliminary results of RBC collaboration using quenched
domain wall QCD with the DBW2 gauge action on an 
$L^3\times T\times N_5=16^3\times 32\times 12$ lattice 
at $1/a=1.23(5)$ GeV: $|\alpha(\ndr,1/a)|=0.006(1)$ GeV$^3$ and 
$|\beta(\ndr,1/a)|=0.007(1)$ GeV$^3$ \cite{rbc_02}, which are smaller by 30\% 
than our values. This is of the order of scaling violation  
that is expected when the physical scale is set by mesonic quantities, but
a quantitative comparison awaits their calculation of the
continuum limit.
It should be noted that the effect of the change of renormalization scale
from $1/a=1.23$ GeV to $2$ GeV is about 3.5\% and negligibly small.

In conclusion, we have studied scaling violation in the proton decay
$\alpha$ and $\beta$ parameters in quenched lattice QCD.
Scaling violation is mild if the physical lattice scale is set
by the nucleon mass, whereas a 30\% systematic errors may arise
from the physical scale, reflecting a part of the quenching error.
Our estimate of $\alpha$ and $\beta$ is larger by about 3 times
than the smallest prediction among various QCD models.
This implies stronger constraints on GUT models. 
We can eliminate a remaining major uncertainty due to the quenched 
approximation by repeating the calculation on 
the full QCD gauge configurations which we already have\cite{full}. 
This should be a next task.

\vspace*{3mm}
This work is supported by Large Scale Simulation Program
No.~98 (FY2003) of High Energy Accelerator Research Organization
(KEK), and in part by the Grants-in-Aid of the Ministry of
Education (Nos. 12740133, 13135204, 13640259, 13640260, 14046202,
14540289, 14740173, 15204015, 15540251, 15540279, 15740165).
N.T. is supported by the JSPS Research Fellowship.



\end{document}